# Deep learning based Chinese text sentiment mining and stock market correlation research

ZHANG Chenrui

（School of Economics and Finance, South China University of Technology, Guangzhou 510006, China）

**Abstract:** We explore how to crawl financial forum data such as stock bars and combine them with deep learning models for sentiment analysis. In this paper, we will use the BERT model to train against the financial corpus and predict the SZSE Component Index, and find that applying the BERT model to the financial corpus through the maximum information coefficient comparison study. The obtained sentiment features will be able to reflect the fluctuations in the stock market and help to improve the prediction accuracy effectively. Meanwhile, this paper combines deep learning with financial text, in further exploring the mechanism of investor sentiment on stock market through deep learning method, which will be beneficial for national regulators and policy departments to develop more reasonable policy guidelines for maintaining the stability of stock market.

## 1    Problem formulation

Trading in the stock market is actually a game between people, and this game is also the research direction of many scholars. In the financial field, the fluctuation of stock price also affects everyone's heart. In the 1970s, the American economist Eugene Fama put forward the "efficient market hypothesis", which holds that in the efficient market, the stock price at any given time fully and accurately reflects all market information. After the 1980s, with the emergence of a large number of market anomalies and irrational trading behavior in the market, the "efficient market hypothesis" has been greatly challenged. According to statistics, the number of A-share investors has reached 189 million, which is a very amazing figure, of which the number of natural persons (retail investors) has reached 82% (data source: 2018 Shanghai Stock Exchange). Individual retail investors have a huge position in the field of investment.

According to the conclusion of He Xingzhi (2021) [1], from the development stage of China's stock market, China's stock market has initially possessed the basic characteristics of weak efficiency market. However, there is a phenomenon that investors use fundamental information to obtain excess profits in the stock market, and the stock market is not a semi strong efficient market. John Maynard Keynes (1936) once pointed out that many economic activities are dominated by "animal spirit", and behavioral finance is also born from this operation. Shiller (2017) pointed out, "with the advancement of research methods and the increase of social media data accumulation, text analysis will become a more powerful field in economics in the next few years. Our goal is to promote research in this direction." In recent years, the research on the relationship between investor sentiment and asset pricing has gradually become a hot topic. The research mainly includes the selection of data, the extraction and analysis of sentiment and the construction of model.

## 2  Literature review and review of research methods

### 2.1  Selection of data

Early studies mainly selected structured data to indirectly reflect investor sentiment, such as market information (opening price, closing price, market volume, etc.), economic indicators (macroeconomic indicators, financial statement data), and technical indicators (simple daily moving average, weighted daily moving average, etc.) With the development of text mining and data analysis techniques in recent years, more and more studies have started to use text-based unstructured data that can directly reflect As text mining and data analysis techniques have improved in recent years, more and more studies have started to use text-based unstructured data that can directly reflect investor sentiment, such as news (financial news, general news, company reports, etc.), social media (Twitter, Facebook, WeChat (Shi Shanchong, 2018)[2] , Weibo, blogs, stock bar forums, etc.). Research using stock market data and technical indicators still dominates due to the ease of data extraction and the accuracy of prediction results, but the trend of using social media data for research is gradually emerging. In the future, combining social media data with stock market data and indicators will be a good research direction (Bustos, 2020).

### 2.2  Sentiment Extraction and Analysis

For structured data, market-based indicators are first used to proxy for sentiment, such as trading volume, closed-end fund discounts, and same-day returns on initial public offerings (IPO). Arguably the most influential measure is the (Baker & Wurgler (2006)) investor sentiment index, which is the main component of the six market-based proxies. The second approach is survey-based. Popular consumer indices include the University of Michigan Consumer Sentiment Index and the UBS/Gallup Investor Optimism Index. For unstructured text data, it is mainly through lexicon-and-rules-based packet-of-words techniques and machine learning methods (e.g., word2vec (Mikolov, 2013))[3]. Investor sentiment is extracted from the text and classified in a positive, neutral, negative, etc. manner. Currently, there is no accepted measure of investor sentiment in academia, and how to better extract and measure investor sentiment still needs to be explored (Tang, 2016)[4].

## 2.3 Sentiment Prediction Model

In recent years, machine learning algorithms have been increasingly applied to predictive models, such as support vector machines (SVM), artificial neural networks (ANN), Bayesian models, and deep learning (mainly CNN, ELM, LSTM, DBN, and other methods) are becoming increasingly popular. Deep learning is a subset of artificial neural networks, but unlike traditional machine learning algorithms, it does not require pre-processing of data and extraction of features. (Kraus & Feuerriegel 2017) found that deep learning algorithms have higher accuracy on traditional neural network algorithms for stock market prediction. Support vector machine methods are still widely used in research using machine learning, but the application of deep learning algorithms will be a hot research topic in stock market prediction for a long time to come (Bustos, 2020).

# 3 Selection of data

## 3.1 Selection of text data

In this paper, we crawled the stock bar data of EastMoney.com[1] Stock Bar (zssz399001) from January 1, 2019 to December 31, 2020 by Python. We compared individual stocks as well as index stock bar data horizontally, and finally chose the stock bar of SZCZ. Compared with the SZSE Component Index and other index stock bar data, its data volume is more adequate and the number of viewers is more representative of the vast majority of stockholders' thoughts. Secondly, the SZCZ includes the stocks of 500 listed companies with a certain scale and quality listed on the Shenzhen Stock Exchange. The SZCZ is also compiled by extracting stocks from various industry sectors, so SZCZ can well reflect the situation of stocks in Shenzhen market. In terms of data processing, this paper follows the criteria of text analysis, removes duplicate data, and removes non-text items such as coded images, tables, HTML tags and emoticons.

---

[1] According to a 2018 analysis by iResearch, EastMoney.com is the top financial website in China. With 78 million hours of active browsing time per month, it accounts for 45% of the market share, higher than the remaining nine companies in the top 10 combined.

**Table 1 Daily comment statistics summary information**

| Date | Reviews | Reads | comments | Source |
|---|---|---|---|---|
| 2020-9-9 | 今天抄底爽歪歪了 | 95 | 0 | http://guba.eastmoney.com/news,zssz39900 |
| 2020-9-9 | 创业板公司亚光科技：股东合计减持25%的股份，这不是在减持，分明是在找人接盘准备 | 793 | 1 | http://guba.eastmoney.com/news,zssh000001,963406436.html |
| 2020-9-9 | 如果我专心玩白银可能都不会输这么惨，去年账户上的8万现在只有2万了。 | 317 | 3 | http://guba.eastmoney.com/news,zssz399001,963400740.html |
| 2020-9-9 | 美哥跌耶，我也跌耶，我比美哥跌得黑耶 | 301 | 3 | http://guba.eastmoney.com/news,zssz399001,963404635.html |
| 2020-9-9 | 炒小、差是投资者最基本的选择权利，关键是违规违法没有，既然规则已制定，如果没有违则涨跌应该交给市场 | 183 | 0 | http://guba.eastmoney.com/news,zssz399001,963402753.html |

　　The statistical information obtained from Table 1 shows that the average comment title length is 22.8 characters per day. As can be seen in the table, the content is distributed in a right-skewed manner. In contrast to traditional (relatively short) comments, some very long messages in the stock bar data are often copied and pasted from other sources, such as news reports and analysis reports. We use a simple process to eliminate these potentially influential outliers, keeping only messages of less than 150 characters. In addition, given the uneven number of daily comments and the need to reduce the impact of different reading volumes on the results, we chose to sort the data by reading volume, selecting the top 50 messages per day from highest to lowest for the study.

**Table 1 Text format of pre-processed stock bar comments**

|  | Mean | S.D | Skewness | Min | Max | Count |
|---|---|---|---|---|---|---|
| Data length (characters) | 22.86383 | 13.40368 | 0.126406 | 2 | 66 | 32298 |

Note: This table provides summary statistics for the length of comments per day. Sample mean, standard deviation (S.D.), maximum and minimum number and total number of comments for each variable. The sample contains data on SZCZ stock bar comments for the sample period January 1, 2019 to December 31, 2020.

### 3.2 Selection of stock trading data

We selects SZCZ as the research object, in which daily trading data are selected for study, including the day's closing price, opening price, high price, low price, yesterday's closing price, up or down amount, up or down, volume, turnover and other data, the specific format of the data is shown in Table 3. In this paper, the data is obtained through the open source Python data API - Tushare[2], and the returned results are of the Pandas.DataFrame data type.

**Table 3 Un-normalized quotation data**

| Date | Close | Open | High | Low | Pre_close | Change | Pct_chg | vol | amount |
|---|---|---|---|---|---|---|---|---|---|
| 20201231 | 14470.68 | 14226.28 | 14476.55 | 14226.28 | 14201.57 | 269.1178 | 1.895 | 3.72E+08 | 5.11E+08 |
| 20201230 | 14201.57 | 13970.45 | 14208.68 | 13968.09 | 13970.21 | 231.3549 | 1.6561 | 3.52E+08 | 4.69E+08 |
| 20201229 | 13970.21 | 14042.79 | 14082.5 | 13915.89 | 14044.1 | -73.89 | -0.5261 | 3.72E+08 | 4.78E+08 |
| 20201228 | 14044.1 | 14020.95 | 14112.59 | 13959.14 | 14017.06 | 27.0435 | 0.1929 | 3.73E+08 | 4.83E+08 |
| 20201225 | 14017.06 | 13879.24 | 14017.06 | 13835.52 | 13915.57 | 101.4832 | 0.7293 | 3.38E+08 | 4.35E+08 |

Source：Tushare,http://tushare.org/.

Since the magnitudes may differ between various data, we need to perform normalization of the data to ensure that the data maintain a consistent distribution across model training. The normalization method we use the outlier normalization (Min-Max Normalization) (Patro et al., 2015)[5]. Primarily, the features are mapped to between [0,1] as follows.

$$X = \frac{x - min(x)}{max(x) - min(x)} \quad (1)$$

---

[2] Tushare is a Python-language based open source financial data interface package, mainly to achieve the financial data such as stocks from data collection, cleaning and processing to data storage process, can provide financial analysts with fast, neat and diverse easy to analyze data.

Table 4 Discrepancy normalized quotation data

| Date | Close | Open | High | Low | Pre_close | Change | Pct_chg | vol | amount |
|---|---|---|---|---|---|---|---|---|---|
| 20201231 | 1.00000 | 1.00000 | 1.00000 | 1.00000 | 1.00000 | 0.83034 | 0.73689 | 0.36816 | 0.44054 |
| 20201230 | 0.96354 | 0.96449 | 0.96310 | 0.96421 | 0.96747 | 0.80357 | 0.71986 | 0.33501 | 0.39241 |
| 20201229 | 0.93220 | 0.97453 | 0.94571 | 0.95698 | 0.97786 | 0.58719 | 0.56437 | 0.36716 | 0.40181 |
| 20201228 | 0.94221 | 0.97150 | 0.94986 | 0.96297 | 0.97406 | 0.65874 | 0.61560 | 0.36985 | 0.40789 |
| 20201225 | 0.93854 | 0.95183 | 0.93670 | 0.94584 | 0.95979 | 0.71151 | 0.65382 | 0.31269 | 0.35237 |

Source：Tushare,http://tushare.org/.

## 4  Sentiment analysis of physical texts

After the text sentiment sense was proposed (Bo Pang, 2002) [6], the early metrics for text sentiment were mainly in the construction of sentiment dictionary method, which assigns the words expressing sentiment that frequently appear in the articles and compiles them into a dictionary, and scores the matches of the articles by using the dictionary. This method is more general, and You, Jing-Yi Wang and Yi-Ping Huang construct sentiment dictionaries in the context of fintech, assign different weights to the sentiment words in the reports according to the frequency of positive and negative sentiment words appearing in the articles, the number of words in the positive and negative sentiment dictionaries, and other indicators, respectively, after which the positive and negative sentiment indices in each report are calculated, and the net sentiment indices of the reports are obtained by direct summation [7].

Compared to the more subjective assignment and classification of sentiment lexicons, the machine learning approach is more objective and has performed better in different areas of text analysis research. The first task of the machine learning method is to construct a corpus. AI-Nasseri and Ali took the news text information of selected companies in forums related to their companies and predicted them using models trained in software using plain Bayesian, decision tree, and support vector machine (SVM) algorithms [8]. Pawar et al. combined recurrent neural networks (RNN) with long and short term memory units (LSTM) to predict the stock market and compared it with the traditional support vector machine. LSTM) combined for stock market prediction and compared with traditional support vector machine and plain Bayesian classifier [9].

## 4.1 Model Design

Natural language processing (NLP) pre-training efforts in the deep learning era make extensive use of Word Embedding. When training with deep learning models, the trained subsets are transformed into word vectors as the input layer of the neural network. In the process of deep learning model training, the degree of good or bad training result depends on the size of the training set, and a larger training set can produce better word vectors. At present, most of the task models in the field of natural language processing use trained word vectors. In the training process of word vectors, word vectors ignore contextual ideation, and when words have multiple meanings, they often correspond to the same word vectors, which is unreasonable; therefore, in 2018, Devlin et al. proposed the pre-trained language model BERT, which was a big step forward in the field of NLP by refreshing the list of 11 NLP tasks as soon as it was introduced [10].The BERT model's structure is shown in Figure 1, where $E_1, E_2, \ldots, E_N$ are the input characters of the model, and the input characters are used to obtain text features through a bidirectional Transformer feature extractor, and the corresponding vectors $T_1, T_2, \ldots, T_N$ are output after the input characters are trained.

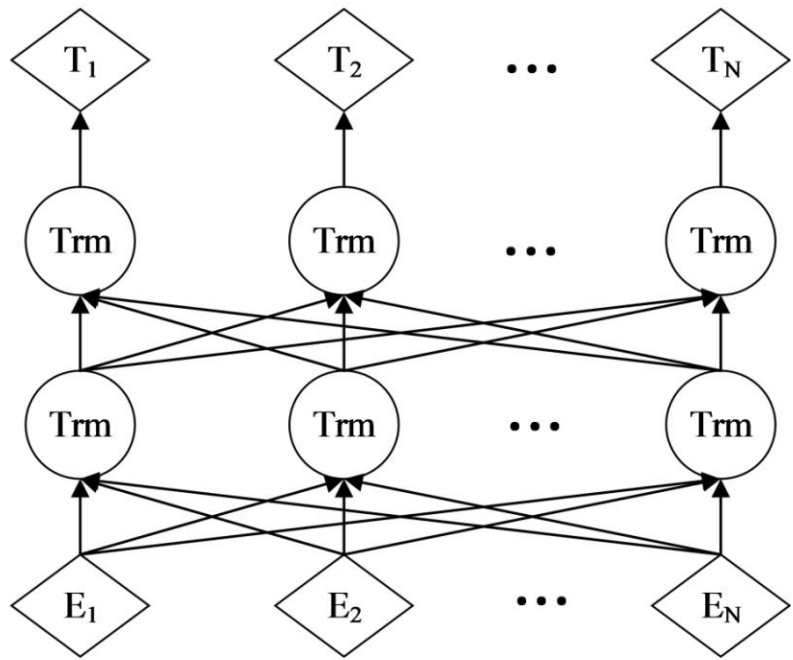

**Figure 1 BERT model structure**

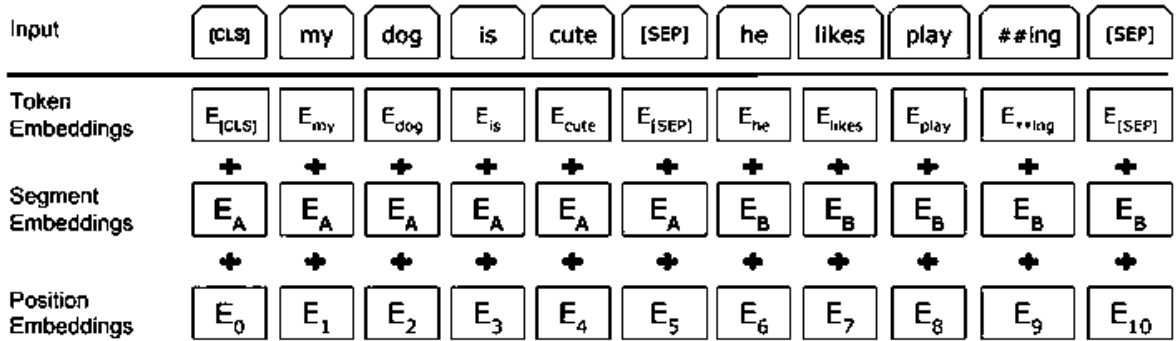

**Figure 2 Training method of BERT model**

As shown in Figure 2, BERT is trained as a model in which all layers are able to incorporate contextual semantics. Its input consists of three vectors of Token Embeddings, Segment Embeddings, and Position Embeddings. Meanwhile, BERT uses the MLM model (Masked Language Model), which masks a part of words, similar to fill-in-the-blank. Then the masked model is predicted to achieve the purpose of contextual training by iteration.

## 4.2 Model training based on financial corpus

BERT is essentially a two-stage NLP model. The first stage is called: Pre-training, where an existing unlabeled corpus is used to train a language model. This phase is very time consuming and requires very high computing power usually requiring 4 to 16 cloud TPUs for more than 4 days. Due to hardware limitations and accuracy requirements, we choose the Chinese BERT pre-training model Chinese-BERT-wwm based on Whole Word Masking (wwm) released by Joint Laboratory of HIT and iFLYTEK Research (HFL) as the pre-training model. The second stage is called Fine-tuning, which uses the pre-trained language model to complete specific NLP downstream tasks. pre-training is expensive, while Fine-tuning is relatively less expensive. It is precisely on local that we take Fine-tuning and use the financial corpus to train the BERT model.

The corpus is input to Encoder module to get the transformed indexes after the word splitting, so as to get the word vector of each word. This is different from the splitting of words by a splitting tool such as jieba, e.g., the phrase "看来进入牛市了，大盘大涨，能带动投资情绪上涨" is split to get: [看，来，进，入，牛，市，了，大，盘，大，涨，能，带，动，投，资，情，绪，上，涨]. Thus, the words are combined with the corpus table in the BERT pre-training model. In the experiments of this paper, the maximum word sequence length is set to 128 bits, and those that are not full 128 bits will be filled using 0. The [CLS] and [SEP] tags are also added at the beginning and end of the sentences. After the transformation of the BERT input sentences is completed, there are two training methods, Masked LM and Next Sentence

Prediction (NSP) next sentence prediction.

### 4.2.1 Masked LM

BERT training uses [MASK] to replace part of the words in a sentence to make the model use the context to make predictions. For example, if the sentence "看来进入牛市了，大盘大涨，能带动投资情绪上涨", there is an 80% probability that the sentence will be changed to "看来进入牛市了，大盘大[MASK]，能带动投资情绪上涨", and the rise in the sentence will be replaced by [ MASK] instead of the sentence, there is a 10% probability of keeping the sentence unchanged, there is also a 10% probability of replacing "up" with other words such as: "看来进入牛市了，大盘大跌，能带动投资情绪上涨". This 8:1:1 replacement strategy is mainly to avoid the subsequent use of the word [MASK], which leads to performance impact.。

### 4.2.2 Next sentence prediction (NSP)

The second task in the BERT training is the next sentence prediction, which is also intended to allow the model to perform the task with contextual semantics under supervised learning. In the same example of "看来进入牛市了，大盘大涨，能带动投资情绪上涨", during the training process, there is a 50% probability of selecting two sentences that are connected: "[CLS] 看来进入牛市了，大盘大涨，能带动投资情绪上涨[SEP]沪深两市翻红[SEP]". There is also a 50% probability that the unrelated sentences are connected: " [CLS] 看来进入牛市了，大盘大涨，能带动投资情绪上涨[SEP]美股受大挫[SEP]", and "No" is output in the label.

### 4.2.3 Fine-tuning using the financial corpus

BERT can be used for the task of sentiment recognition of financial entities after completing pre-training. In sentiment analysis [CLS] will be used as the output of the next network. According to the specificity of financial texts, Fine-tuning using financial texts with sentiment annotation will allow training models with higher accuracy in specific domains such as finance.

## 5 Experiment and Analysis

### 5.1 Experimental environment

The experimental environment is shown in Table 5.

**Table 5 Experimental environment**

| Development Environment | Parameters |
|---|---|
| CPU | R7-3750H(2.30GHz) |
| GPU | GTX1660Ti 6GB |
| RAM | 16GB |
| System | Windows 10 64-bit |
| Framework | TensorFlow |
| Tools | PyCharm |

### 5.2 Dataset

The selected data from January 1, 2019 to December 30, 2019 were selected for the annotation of artificial sentiments. In this paper, the sentiments are labeled using triple classification as shown in Table 6, where 0 denotes negative sentiment, 1 denotes neutral sentiment, and 2 denotes positive sentiment, and the text is divided into training set and test set in the ratio of 8:2 for training.

**Table 6 Symbolic representation of sentiment classification**

| Negative | Negative | Negative |
|---|---|---|
| 0 | 1 | 2 |

### 5.3 Parameter Setting

The model used the pre-trained model of english_roberta_wwm_large_ext_L-24_H-1024_A-16 (24-layer, 1024-hidden, 16-heads) released by Joint Laboratory of HIT and iFLYTEK Research (HFL). The total size of the parametric model is 330 MB. the batch size is 16, the learning rate is 2e-5, and the max seq length is 128.

## 5.4 Model training results

The training result is an accuracy of 0.7553 and a loss of 0.6558. A partial sample of the prediction results is shown in Table 7. As Figure 3 shows the prediction of the review data from January 1, 2020 to December 31, 2020 based on the trained model, it can be seen that there are mostly negative sentiments and few neutral sentiments.

**Table 7 Some sample data**

| Date | Reviews | Sentiment | Positive_Pro | Negtive_Pro |
| --- | --- | --- | --- | --- |
| 2020-6-9 | 北上今天净流入60亿，尾盘猛进二十亿，明天大盘无忧！ | 2 | 0.938666 | 0.0613335 |
| 2020-6-9 | 大跌正式开始 | 0 | 0.00298048 | 0.99702 |
| 2020-6-9 | 三家财务造假，暴风集团，东方金钰，长城影视，股价跌停 | 0 | 0.0382706 | 0.961729 |

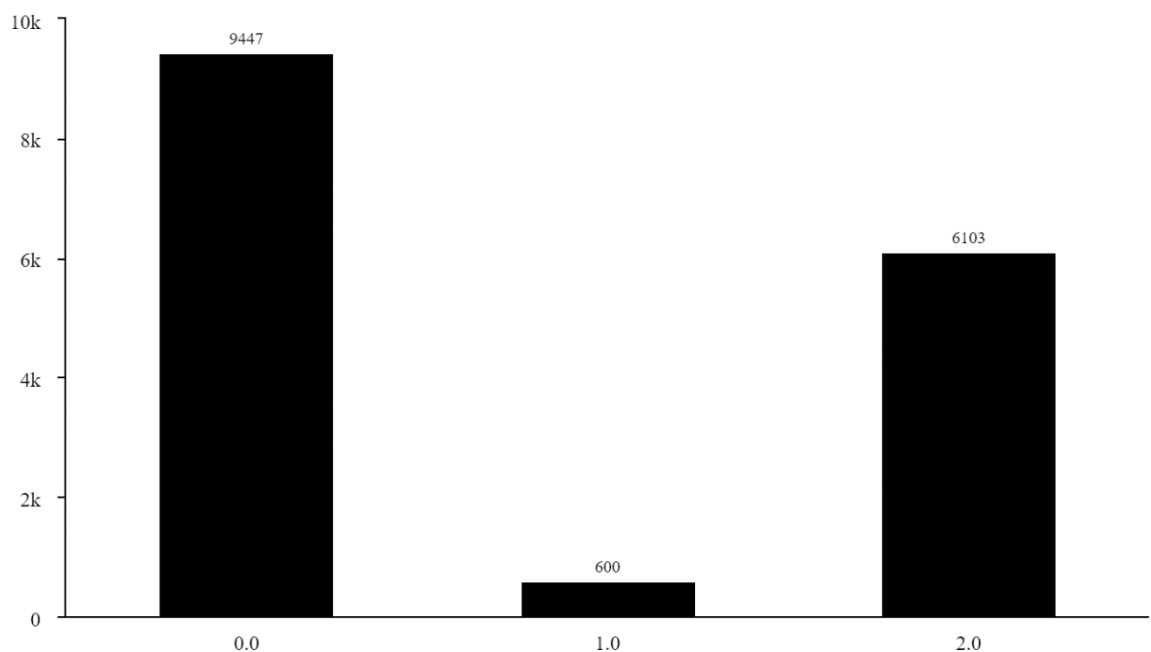

**Figure 3 Data distribution in 2020**

## 5.5 Empirical studies

### 5.5.1 Calculation of correlation of emotional data

For data with multiple entries per day, the following formula for processing data by date.：

$$sentiment_t = P_{positive} + (-1) * P_{negative} \quad (2)$$

$$emotions^T = \frac{\sum_{t=1}^{n} sentiment_t}{n} \quad (3)$$

The $sentiment_t$ denotes the sentiment score of each comment. The $P_{positive}$, $P_{negative}$ denote the probability that the sentiment is positive or negative, respectively. The sentiment index $emotions^T$ denotes the average of all sentiment scores during day T. $emotions^T \in (0,1)$. If $emotions^T$ converges to 0, the market sentiment is negative. If $emotions^T$ tends to 1 then market sentiment is positive.

Also, since the magnitudes of the data may be different, we need to normalize the data to ensure a consistent distribution across model training. Therefore, the data of this experiment are normalized using formula (1).

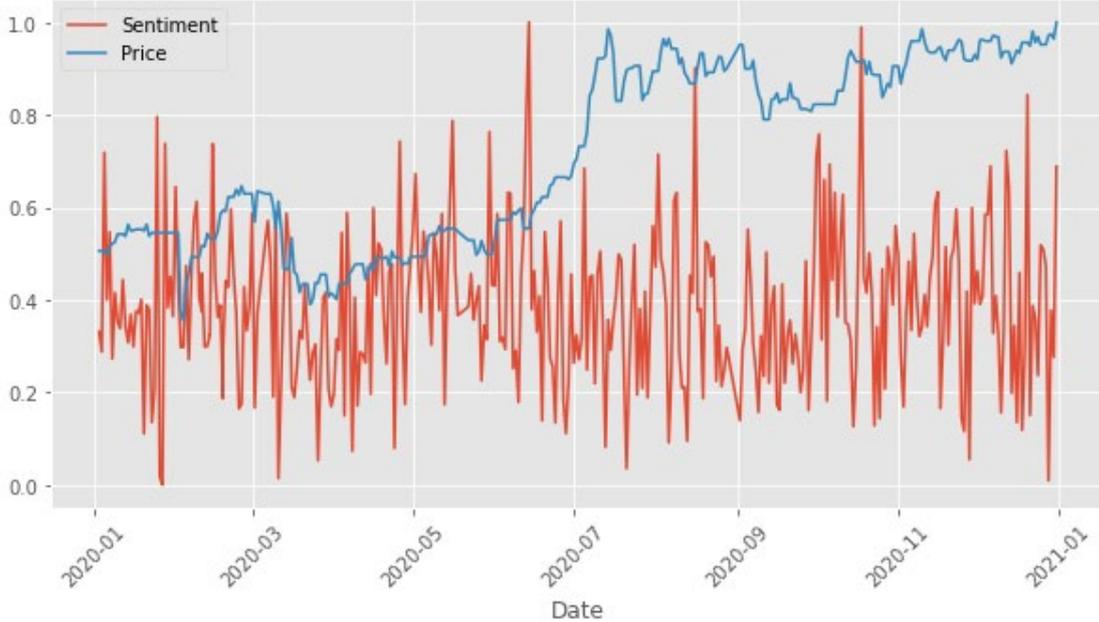

**Figure 4 Market sentiment and stock price normalization results**

As can be seen from Figure 4, the relationship between the sentiment index and the stock price is not very obvious. This experiment follows a time window size of 30, each window contains a minimum number of observations of 1 and smoothing the data of variables Sentiment and Price. The 30-day averages of Sentiment and Price variables are derived and the calculated avg_Sentimen and avg_Price replace the original Sentiment and Price. the results are shown in Figure 5, in that the market sentiment shows homogeneous fluctuations with stock prices.

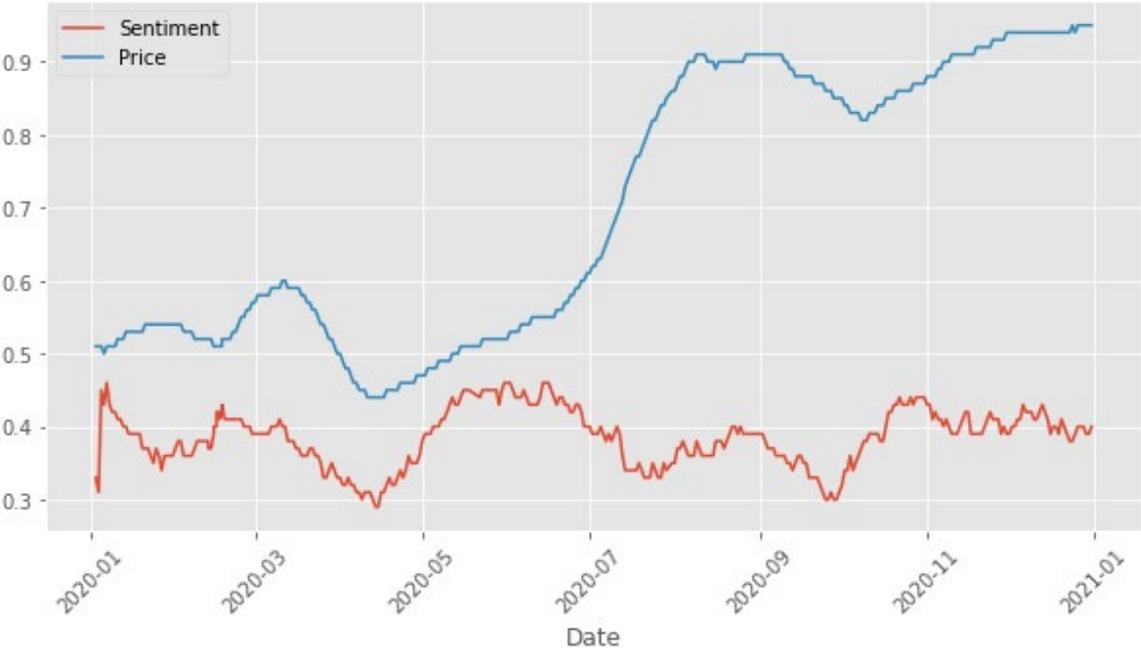

**Figure 5 Market sentiment and stock price smoothing results**

# 6 Research the correlation of variables based on the MIC

## 6.1 Theory of maximum information coefficient

In 2011, David N. Reshef et al. of Harvard University [11] proposed the maximum information coefficient (MIC).MIC is a good measure of the interdependence between variables, and it has two important properties: extensiveness and fairness. The extensiveness of MIC means that it is Multi-sample case is sensitive to a wide range of functional relationships, and can detect a wide range of relationship types, such as non-functional relationships and hyperfunctional relationships synthesized from multiple functional relationships, etc. The fairness of MIC means that when the same noise is added to different relationship types, the MIC values between them are similar. Conversely, when the MIC values of two variables are calculated to be similar or equal, the values are also similar for the degree of noise added.

## 6.2 Processing of MIC[12]

Given a finite ordered data set X = {x1,x2,x3,···,xn}. If the x-axis is divided into x grids and the y-axis is divided into y grids, then an x × y grid division G is obtained, where x, y are positive integers, and the number of points falling into G as a proportion of the number of X is considered as its probability density X|G. And the probability distribution X|G is obtained differently according to different grid divisions. In X = {x1,x2,x3,···,xn}, the mutual information between two variables xi and xj can be defined as follows.

$$I(x_i, x_j) = \sum_{x_i \in X} \sum_{x_j \in X} p(x_i, x_j) \log_2 \left( \frac{p(x_i, x_j)}{p(x_i) p(x_j)} \right) \qquad (4)$$

Given x, y, if the values of x, y are changed, the obtained mutual information values are also changed and the largest of them is recorded as I (x,xi,xj). Then normalization is performed to compare the data sets in different dimensions and the normalized values are between [0,1]. By changing the values of x, y, the feature matrix of the normalized mutual information values between the variables can be obtained. The maximum value of the eigenmatrix is the value of the maximum information coefficient MIC between the two variables.

$$M(X)_{x_i, x_j} = \frac{I(X, x_i, x_j)}{\log(\min\{x_i, x_j\})} \qquad (5)$$

The set X={x1,x2,x3,⋯,xn} sample capacity takes the value of n, and the fraction of latticework takes the value less than B(n). Then the maximum information coefficient can be defined as follows.

$$MIC(X) = \max_{xy<B(n)} \{M(X)_{x,y}\} \qquad (6)$$

In the above equation, x,y is the number of divided lattices in the x-axis y-axis direction, that is, the grid distribution, where B(n) is a variable, and the size of B(n) is generally about 0.6 times the data n, that is, B(n) ≈ n0.6.

### 6.3 Calculation of MIC

In Python we can do the MIC calculation with the help of minepy library. In this paper, avg_Sentimen and avg_Price are used as variables to calculate the MIC of both, and the results are shown in Figure 6.

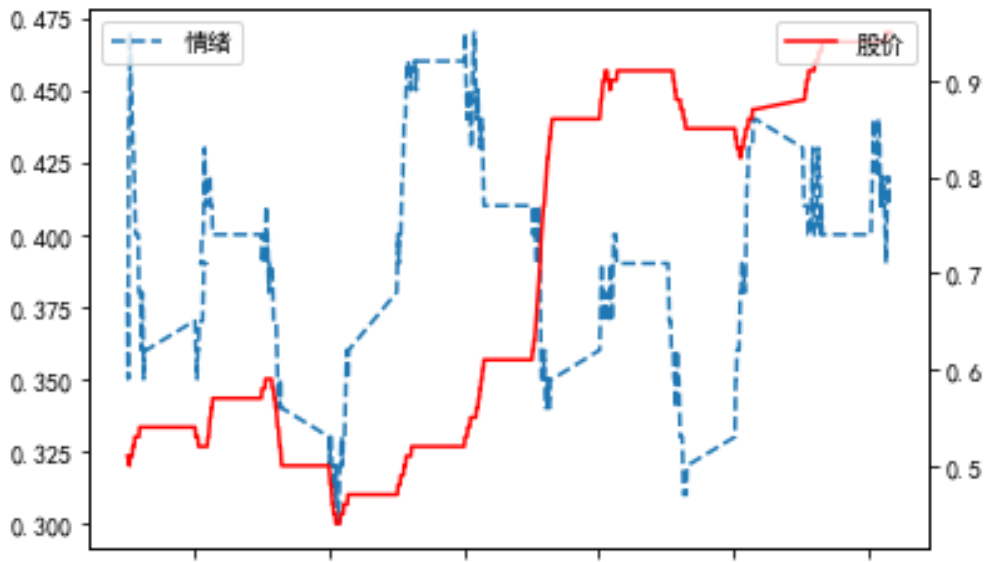

**Figure 6 Calculation results of avg_Sentimen and avg_Price**

$MIC$ (avg_Sentimen, avg_Price) = 0.3806609398310775

The value of MIC is about 0.38, indicating that avg_Sentimen and avg_Price show a certain degree of correlation.

# 7 Summary

The continuous development of deep learning provides a good framework for dealing with financial problems. This paper also shows the effectiveness of deep learning models in the field of finance. The BERT model built out based on this is a good solution to the problem of multiple meanings of words in deep learning such as Word2Vec. The BERT model trained based on financial corpus has a better performance in stock prediction, and it can better improve the accuracy of prediction by combining sentiment analysis based on the traditional method for stocks. In addition, the construction of quantitative indicators of investor sentiment and the study of investor sentiment affecting the stock market are beneficial for investors to grasp the trend in investment and obtain excess returns according to it. On the other hand, at the same time, since there are a large number of retail investors in China's stock market, investor sentiment changes have a certain influence on the stability of the stock market, and further exploring the mechanism of investor sentiment's influence on the stock market through deep learning methods will be beneficial to national regulators and policy departments in formulating more reasonable policy guidelines for maintaining the stability of the stock market.

# 8 Shortcomings and future prospects

For the selection of financial text, this paper selects stock bar as the object of study. During the study, it was found that the stock bar has very high data noise due to its spontaneity, which is not conducive to the fitting of data and will have an impact on the effect of deep learning model. In the later stage, we hope to use financial news text data for the study. On the one hand, financial news text data has a unified standard, on the other hand, it can directly feedback the actual situation of listed companies, which is easy to fit with the market. In addition, due to the limitations of the author's research conditions and machine performance, the deep learning model cannot achieve the optimal training effect in the training process, and later on, we can consider adopting cloud computing and other methods to conduct research with sufficient financial support.

# References

[1]. 贺行知.上海股票市场有效性假说研究[J].投资与创业,2021,32(20):79-82.

[2]. 石善冲,朱颖楠,赵志刚,康凯立,熊熊.基于微信文本挖掘的投资者情绪与股票市场表现[J].系统工程理论与实践,2018,38(06):1404-1412.

[3]. Mikolov T.,Sutskever I.,Chen K.,Corrado G.S.,Dean J.,2013,Distributed Representations of Words and Phrases and Their Compositionality [C],Proceedings of Advances in Neural Information Processing Systems.

[4]. 唐国豪、姜富伟、张定胜：《金融市场文本情绪研究进展》[J],《经济学动态》2016 年第 11 期。

[5]. Patro S.,Sahu K.K.,2015,Normalization:A Preprocessing Stage [R],arXiv,No.1503.06462.

[6]. Pang B, Lee L. A sentimental education: Sentiment analysis using subjectivity summarization based on minimum cuts. In: Scott D, ed. Proc. of the ACL 2004. Morristown: ACL, 2004. 271-278.

[7]. 王靖一,黄益平.金融科技媒体情绪的刻画与对网贷市场的影响[J].经济学(季刊),2018,17(04):1623-1650.DOI:10.13821/j.cnki.ceq.2018.03.15.

[8]. Al-Nasseri A,Ali F M.What Does Investors' Online Divergence of Opinion Tell Us About Stock Returns and Trading Volume?[J].Journal of Business Research,2018,86(1).

[9]. Pawar K,Jalem R S,Tiwari V.Stock Market Price Prediction Using LSTM RNN[M]//
Emerging Trends in Expert Applications and Security.Singapore:Springer,2019:493-503.

[10]. DEVLIN J, CHANG M W, LEE K, et al. Bert: pre-training of deep bidirectional transformers for language understanding[C]//Proceedings of the 2019 Conference of the North American Chapter of the Association for ComputationalLinguistics:Human Language Technologies. Stroudsburg, PA: Association for Computational Linguistics, 2019: 4171-4186.

[11]. David N. Reshef et al. Detecting Novel Associations in Large Data Sets[J]. Science(S0036-8075), 2011, 334(6062) : 1518-1524.

[12]. 李明媚,文成林,胡绍林.一种基于最大信息系数预处理的 k-modes 聚类方法[J/OL].系统仿真学报:1-10[2021-12-30].
https://kns-cnki-net.webvpn.scut.edu.cn/kcms/detail/11.3092.V.20210826.1453.013.html.

[13]. 赵宏,傅兆阳,赵凡.基于 BERT 和层次化 Attention 的微博情感分析研究[J/OL].计算机工程与应用:1-8[2021-12-30].
https://kns-cnki-net.webvpn.scut.edu.cn/kcms/detail/11.2127.TP.20211115.1831.019.html.

[14]. 欧阳资生,李虹宣.网络舆情对金融市场的影响研究:一个文献综述[J].统计与信息论坛,2019,34(11):122-128.

[15]. 李斌、邵新月、李玥阳：《机器学习驱动的基本面量化投资研究》[J],《中国工业经济》2019 年第 8 期。